\documentclass[11pt]{article}
\usepackage{epsf}
\usepackage{axodraw}
\begin{document}
%
\def\doublespaced{\baselineskip=\normalbaselineskip\multiply
    \baselineskip by 200\divide\baselineskip by 100}
\def\singleandabitspaced{\baselineskip=\normalbaselineskip\multiply
    \baselineskip by 120\divide\baselineskip by 100}
\def\singlespaced{\baselineskip=\normalbaselineskip}
\def\mol{Mol}
\def\etmiss{E\llap/_T}
\def\eslt{E\llap/_T}
\def\esl{E\llap/}
\def\msl{m\llap/}
\def\to{\rightarrow}
\def\te{\tilde e}
\def\tmu{\tilde\mu}
\def\ttau{\tilde\tau}
\def\tl{\tilde\ell}
\def\ttau{\tilde \tau}
\def\tg{\tilde g}
\def\tnu{\tilde\nu}
\def\tell{\tilde\ell}
\def\tq{\tilde q}
\def\tu{\tilde u}
\def\tc{\tilde c}
\def\tb{\tilde b}
\def\tst{\tilde t}
\def\tt{\tilde t}
\def\tc{\widetilde C}
\def\tn{\widetilde N}

\hyphenation{mssm}
%
%
%
\begin{titlepage}
\begin{flushright}
UM-TH-98-07\\ 
March 19,1998\\
\end{flushright}

\vspace{0.8cm}

\begin{center}
\mbox{\Large \textbf{MEASURING THE SUPERSYMMETRY LAGRANGIAN}} \\
\vspace*{1.6cm}
{\large Michal Brhlik and G.L. Kane} \\
\vspace*{0.5cm}
{\it Randall Physics Laboratory, University of Michigan} \\
\vspace*{0.00cm}
{\it Ann Arbor, MI~~48109-1120} \\

\begin{abstract}
The parameters of the supersymmetry Lagrangian are the place where experiment 
and theory 
will meet. We show that measuring them is harder than has been thought, 
particularly because of large unavoidable dependences on phases. Measurements 
are only 
guaranteed if a lepton collider with a polarized beam and sufficient energy  to 
produce the relevant sparticles is available. Current limits on superpartner 
masses, WIMPs, and the supersymmetric Higgs are not general, and need 
re-evaluation. 
We also tentatively define the MRM (Minimum Reasonable Model), whose 
parameters may be measurable at LEP, FNAL and LHC. 
\end{abstract}

\medskip
\end{center}
\end{titlepage}

\singleandabitspaced
\section{Introduction}

Supersymmetry is a complete quantum field theory. The form of the full 
Lagrangian is known, including the general effects of ``soft'' supersymmetry 
breaking (all terms which do not introduce power contributions from higher 
scales to lower scales) \cite{dige}, regardless of how supersymmetry is broken. 
Many people have studied possible direct and indirect 
SUSY effects based on this Lagrangian, often by making many assumptions and 
studying the variation of a few parameters.

The full Lagrangian \cite{disu,hab} depends on 107 parameters beyond the 19 of 
the Standard 
Model (SM) of particle physics (including the gravitino). That may seem like a 
large number. But older 
readers may recall that at a similar stage in the development of the SM the 
situation was not so different; for example, if one counts by assuming that one 
knew there were three  quark and three lepton families, allowing for Majorana 
neutrino masses, but that instead of knowing the interaction was V-A it 
could be any of S,V,T,A,P, then there are about sixty parameters in the SM. 
Data gathered over three decades led to the present form. Also, there are 
already many constraints \cite{mas} on the SUSY parameter space from the 
absence of flavor 
changing neutral processes, electric dipole moments, certain collider signals, 
and so on. Most processes only depend on a few parameters, so in practice using 
the general parameter set is not easy but is doable. 

The purpose of this paper is to emphasize that once there is data on 
superpartners, from colliders or rare decays or WIMP signals, it is essential 
to measure the parameters of the Lagrangian without making any of the 
simplifying assumptions that have become standard. One might hope that 
measurements would be insensitive to ignoring some phases, or to various 
degeneracy assumptions. We will give examples below to show that is generally 
not so. 
Indeed, limits on on sparticle masses from colliders, or WIMPs, and the Higgs 
boson mass are sensitive to the assumptions and need 
to be re-evaluated.

It is extremely important to measure the parameters. As understanding of string 
theories improves models will emerge \cite{iba} that predict the 107 
parameters, and the 
measured values will provide perhaps the best tests of theoretical predictions 
and ideas. Even more likely, the historical path will be followed --- once the 
parameters are measured, their pattern will suggest  
how SUSY is broken and how to find the correct vacuum, just as 
gauge invariance was suggested by data as the SM emerged. Unfortunately, the 
relations connecting the data and the parameters are very non-linear, and 
non-physical solutions can be obtained if some parameter is assumed to have a 
value different from its true one. It is important to understand that in almost
all cases the physical observed particle masses are not the soft-breaking 
parameters that appear in the Lagrangian and have a close connection to the 
theory; the particle masses result from diagonalizing mass matrices. Unless a 
theory predicts all the relevant phases and additional soft parameters it 
cannot predict masses of mass eigenstates. Unless the soft  parameters of the 
SUSY Lagrangian can be measured, the essential interplay between experiment and 
theory needed for progress will be threatened.

For this paper we will mainly consider\footnote{The full Lagrangian can be seen 
in many places, e.g. in Ref. \cite{disu,hab}.} the neutralino and chargino 
sectors. The relevant terms in the Lagrangian are
\begin{equation}
{\cal L}=-\frac{1}{2} (M_1 \tilde{B} \tilde{B}+
M_2 \tilde{W} \tilde{W}+M_3 \tilde{g} \tilde{g})+\mu H_{u} H_{d} + c.c.
\end{equation} 
Although $\mu$ arises from the superpotential we will speak of it as soft too 
since its value is of the same order as the other soft tems $M_1,M_2,M_3,...$.
$M_1,M_2,M_3$ are complex (as is $\mu$) but a global symmetry can be 
used to choose one of them real; here we will take $M_2$ real. From now on we 
write $M_1 e^{i \varphi_1}$ and $\mu e^{i \varphi_{\mu}}$ with $M_1$ and $\mu$ 
real. 
The other important parameter we need is the ratio of the vacuum expectation 
values of the two Higgs doublets, $\tan \beta$, which we choose to be real by 
setting the phase of the $B\mu$ soft parameter to zero using a second
global symmetry. Some earlier work on relating theory and experiment once 
superpartner data is available, and on measuring parameters in the chargino and 
neutralino sector, can be found in Ref. \cite{feng}.

In the next section we consider the simplest sector, the two chargino states 
(linear combinations of wino and charged higgsino states). Then we examine the 
neutralino sector alone and in combination with charginos, including the 
effects relevant to the LSP as cold dark matter (CDM). Then we look at 
implications for the Higgs sector, and finally at the extension of these 
considerations to the full MSSM.

\section{The Chargino Sector}

The chargino mass matrix is 
\begin{equation}
{\cal M}_C=\left(\begin{array}{cc}
 M_2 &  \sqrt{2} M_W \sin\beta   \\
\sqrt{2} M_W \cos\beta   & \mu e^{i\varphi_{\mu}}  \end{array}\right),
\end{equation}
so it depends on $M_2$, $\tan\beta$, $\mu$, $\varphi_{\mu}$. The two chargino 
masses $m_1$ and $m_2$ are the eigenvalues of ${\cal M}_C^\dag{\cal M}_C$. They 
are a little complicated, so consider   
\begin{equation}
Tr({\cal M}_C^\dag{\cal M}_C)=m_1^2+m_2^2=M_2^2+\mu^2+2 M_W^2 ,
\end{equation}
\begin{equation}
Det({\cal M}_C^\dag{\cal M}_C)=m_1^2 m_2^2=M_2^2 \mu^2+2 M_W^4 \sin^2 2\beta
-2 M_W^2 M_2 \mu \sin 2\beta \cos\varphi_{\mu} .
\end{equation}
One can immediately see from eq. 2.3 that the value of $\tan\beta$ extracted 
from data on chargino masses is very sensitive to $\cos\varphi_{\mu}$, so if 
different values for $\varphi_{\mu}$ are assumed, different values for 
$\tan\beta$ 
will result.

One might argue that $\varphi_{\mu}$ is constrained by the electron or neutron 
electric dipole moment and may be small, but that is not clear. 
First, in other sectors additional phases enter and our use of $\varphi_{\mu}$ 
should be considered mainly as an example of the role of phases, demonstrating 
how 
they can be important. Second, even the situation for $\varphi_{\mu}$ itself is 
more complicated. No symmetry is known that implies it should be small. The 
analysis of the electric dipole moments has not been done with a general phase 
structure, and constraints will surely be smaller than when only 
$\varphi_{\mu}$ is included. For example, Nath and Ibrahim \cite{nath} have 
included 
two phases and concluded that the constraint from $d_n$ on $\varphi_{\mu}$ 
could be rather weak. It may 
finally turn out that there is indeed a constraint, but that has yet to be 
established. Recently Garisto and Wells have examined how some of the phases 
evolve from high to low scales \cite{gari}. 

Thus we see that from just the chargino masses two observables are available 
for four parameters. The phase in eq. 2.1 cannot be rotated away by rotating 
the wino and higgsino fields. When one adds production cross sections and 
asymmetries 
the situation does not improve since there is a sneutrino exchange diagram; 
then additional parameters enter and no net information is obtained 
(we consider the polarized beam case below). Thus 
measurements at LEP or FNAL or LHC of the chargino sector alone cannot 
determine 
$\tan\beta$ or $\mu$ or $\varphi_{\mu}$ or $M_2$. To measure the parameters it 
is necessary to have data with polarization of a beam or final particles; for 
the latter case analyzing the polarization cannot be allowed to depend on 
assumptions about the 
relative size of various decay branching ratios, and requires very large 
statistics \cite{djou}. In particular, if charginos decay via a spinless 
slepton, as is somewhat likely, their polarization will not be transmitted to 
the final 
lepton. Again we stress that even if 
it should turn out that $\varphi_{\mu}$ is  not large, here we are emphasizing 
that dependence on many phases can be significant. Further, even if 
$\varphi_{\mu}$
is small (or $\pi-\varphi_{\mu}$ is small), there are still three parameters in 
the chargino mass matrix so at least three observables depending only on those 
parameters must be measured.

\section{The Neutralino Sector}

Next consider the neutralino sector. Its mass matrix in the MSSM is $4\times 
4$, with additional parameters $M_1$ and the associated phase $\varphi_1$. 
First 
we can ask if adding masses and cross sections of neutralinos can allow a 
measurement of any parameters. In this case there are six parameters
($M_1$, $\varphi_{1}$, $M_2$, $\tan\beta$, $\mu$, $\varphi_{\mu}$) for the 
combined 
chargino + neutralino sectors, so 
measurements of six masses could allow their determination, although several 
solutions are likely to exist since all of the relationships of parameters to 
data are non-linear and complicated. At LEP it is no longer possible to produce 
all 
six states, but at FNAL it may be possible when the upgraded collider takes 
data.

 However, it is not clear that the needed masses can be measured at FNAL 
(or LHC) even if the states are produced. Again, cross sections are not helpful 
since diagrams with slepton or squark exchange can be large. Below we will see 
that at a 
lepton collider with a polarized beam (the Next Polarized Lepton Collider, 
NPLC) the situation can be much better. Also, there exists a set of assumptions 
that may permit a meaningful determination of parameters at FNAL (Section 
VII). Figure 1 shows that neutralino masses can depend sensitively on 
$\varphi_{1}$; they can vary even more for other values of parameters and they 
depend strongly on $\varphi_{\mu}$ as well.
\begin{figure}[t]               
\centering
\epsfxsize=4.0in                
\hspace*{0in}
\epsffile{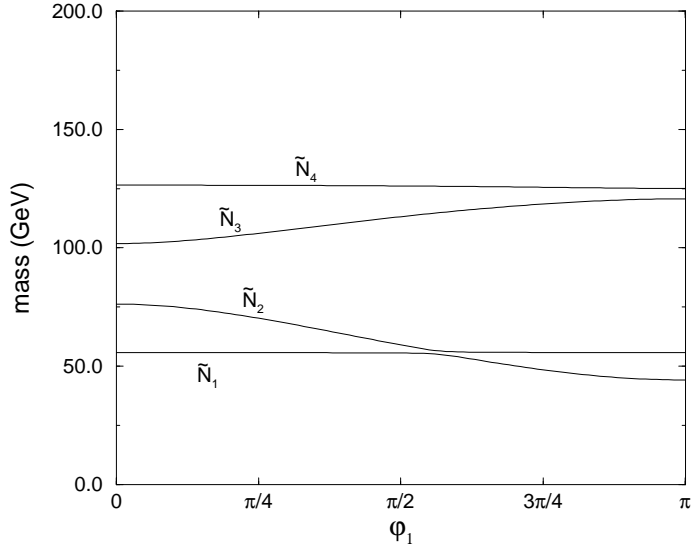}
\caption{ This figure shows the variation of the neutralino mass eigenstates 
with 
$\varphi_1$ for fixed values of other parameters (this graph is for $\mu=-56$ 
GeV,
$\varphi_{\mu}=0$, $M_1=75$ GeV, $M_2=82$ GeV, $\tan\beta=1.2$; for other 
values there is more or less variation). $\tn_1$ is the LSP, a good candidate 
for cold dark matter. We know of no restrictions on $\varphi_1$, though 
when electric dipole moment limits are analyzed with $\varphi_1$ included some 
restrictions may occur. Since the LSP mass varies (and couplings too),  
efficiencies for sparticle searches will vary with $\varphi_1$ (and thus 
limits), as will the resulting   $\Omega h^2$ (and limits) for WIMP 
scattering. (This figure is only meant to illustrate the variation for a 
typical case and not necessarily to be a serious model.) }
\label{fig1}
\end{figure}

\section{The Higgs sector}

The Higgs sector potential depends on the general form of the soft breaking 
parameters. As noted before, it is convenient to set the phase of the $B\mu$ 
soft parameter to zero. Then the tree level Higgs potential is not affected by 
any phases since $m_{H_1}^2$ and $m_{H_2}^2$ are real and $\mu$ enters only 
through $|\mu|^2$. Minimization conditions for the tree level potential in this 
convention do not differ from the phaseless case and $\tan\beta$ is real. The 
Higgs masses and the mixing angle $\alpha$ also do not change at this point and 
therefore the $hZ$ production cross section does not feel the presence of 
complex phases. 

However, the phases do enter through the one-loop correction to 
the effective potential. As an example we consider only the contribution from 
the
top-stop sector with the potential $V=V_{tree}+\Delta V_1$, where 
\begin{equation}
\Delta V_1=\frac{3}{32}\bigg[\sum_{\tilde{t}_1,\tilde{t}_2} m_{\tilde{q}}^4
\bigg(log\frac{m_{\tilde{q}}^2}{Q^2}-\frac{3}{2}\bigg)-2 m_t^4 
\bigg(log\frac{m_t^2}{Q^2}-\frac{3}{2}\bigg)\bigg].
\end{equation}
The phases affect the 
eigenvalues of the stop mass matrix so the 
Higgs potential also changes. This can lead to changes as much as 25 
percent in the light Higgs mass. The effect is largest when $A_t$ and 
$\mu\cot\beta$ are comparable in size; the relative phase of $A_t$and $\mu$ is 
what matters.

\section{WIMPs}     

The lightest eigenvalue (LSP) of the neutralino mass matrix is a candidate to 
provide some or all of the cold dark matter (CDM) of the universe. Once 
superpartners are detected at colliders, and the soft parameters and 
$\tan\beta$ measured, it will be possible to calculate the contribution of the 
LSP to CDM. The first question will be whether the LSP CDM gives a major 
contribution to $\Omega h^2$. Even if superpartners are detected at LEP there 
may not be enough information to answer this question quantitatively.
From FNAL data it may 
be possible to give a tentative answer, and from NPLC a definitive answer can 
be provided if 
the energy of NPLC is large enough to produce all the charginos and 
neutralinos. The second stage is to ask  whether $\Omega h^2$ can be calculated 
to a few percent accuracy; that appears to be possible from NPLC information
\cite{bk}.

These statements follow because the calculation of $\Omega h^2$ requires a 
knowledge of the elements of the neutralino mass matrix, all of which enter 
into calculating the neutralino annihilation cross section that determines the 
neutralino relic density. Thus the calculation can only be done reliably after 
($M_1$, $\varphi_{1}$, $M_2$, $\tan\beta$, $\mu$, $\varphi_{\mu}$) are 
measured. $\Omega h^2$ can vary by a factor of a few depending on the phases.

Similarly, it will be very exciting when a signal for a WISP (Weakly 
Interacting Supersymmetric Particle) is detected directly  in the laboratory.
Until a signal is observed it is useful to set limits on WISPs; the limits are 
sensitive to the phases. For example, the event rate on Ge can vary by nearly 
an order of magnitude as $\varphi_1$ varies. Present limits need to be 
re-evaluated.

Falk, Olive, and Srednicki showed \cite{falk} that whenever sfermion mixing was 
large ( therefore, at least for annihilations involving the stop sector), and 
even more so if the the phase of $A_t$ were large, WISP limits had to be 
re-evaluated. That is an example of our general argument, though the main 
results we 
discuss here from the chargino and neutralino sector occur in addition to this 
effect and are not related to it; our effects are large even for WISPs for 
which sfermion mixing has a negligible effect on annihilation or scattering. 

\section{The Next Polarized Lepton Collider (NPLC)}

At NPLC the key technique is to produce charginos with a polarized beam. Then 
the sneutrino exchange diagram is absent for a right-handed polarized electron 
beam, and one can add cross sections and forward-backward asymmetries as 
observables for the two chargino states $\tilde{C}_{1,2}$. There are then 
twelve possible chargino and neutralino observables, the two chargino masses, 
the four neutralino masses, the cross sections for production of 
$\tilde{C}_{1}+\tilde{C}_{1}$, $\tilde{C}_{1}+\tilde{C}_{2}$,
 $\tilde{C}_{2}+\tilde{C}_{2}$ ($\sigma_{11}$,
$\sigma_{12}$, $\sigma_{22}$), and the associated asymmetries $A_{11}^{FB}$, 
$A_{12}^{FB}$, $A_{22}^{FB}$.  
Measurement of seven or eight should suffice to determine the six parameters 
$M_1$, $\varphi_1$, $M_2$, $\mu$, $\varphi_{\mu}$, $\tan\beta$. Thus 
$\tan\beta$ and other parameters can be measured at NLPC. If all twelve 
can be 
measured the resulting accuracy may be quite good. (We are presently studying 
models to understand how well the parameters can be measured.) Once this is 
done, the LSP relic abundance can be calculated to interesting accuracy 
\cite{bk}. The values of these parameters will be important inputs to measuring 
most of the soft parameters from FNAL and LHC data. The polarization effects 
will also help separate slepton soft parameters, including phases. These 
results imply that a 
lepton collider with a polarized beam and sufficient energy is a necessary 
condition for a complete measurement of the parameters of the supersymmetry 
Lagrangian, a stronger result than has been previously available.

\section{Sleptons, Squarks and the MRM}

Since it will be a long time before there is sufficient data to measure all of 
the MSSM parameters, it is desirable to make some assumptions that reduce the 
number. But it is important to do so in a way that retains 
sufficient generality to have a good chance of correctly describing nature. 
From the gaugino sector we have  a minimal set $M_1$, $\varphi_1$, $M_2$, 
$M_3$, $\varphi_3$, $\mu$, $\varphi_{\mu}$, $\tan\beta$.

If flavor mixing is small, a reasonable assumption, then we can take the 
slepton and squark mass matrices as diagonal. From hermiticity it follows that 
the diagonal elements are real. Similarly we can assume the soft Yukawa 
coefficients $A_k$ are flavor-diagonal; the $A$'s will still be complex in 
general. At this stage the slepton parameters are $m_{{\tilde{e}}_L}^2$,  
$m_{{\tilde{\mu}}_L}^2$, $m_{{\tilde{\tau}}_L}^2$, $m_{{\tilde{e}}_R}^2$, 
$m_{{\tilde{\mu}}_R}^2$, $m_{{\tilde{\tau}}_R}^2$, $A_{e}$, $A_{\mu}$, 
$A_{\tau}$, $\varphi_{A_{e}}$, $\varphi_{A_{\mu}}$, $\varphi_{A_{\tau}}$. We 
can go to a minimal reasonable set by taking 
$m_{{\tilde{e}}_L}^2=m_{{\tilde{\mu}}_L}^2=m_{{\tilde{\tau}}_L}^2$, and
$m_{{\tilde{e}}_R}^2=m_{{\tilde{\mu}}_R}^2=m_{{\tilde{\tau}}_R}^2$, and 
$A_{e}=A_{\mu}=0$ (they actually need not be zero but we assume 
their effect to be proportional to the $e$ and $\mu$ masses so they can be 
neglected). The assumption of universality for the soft slepton mass parameters
is somewhat ad hoc at this point since there is no known physical reason why it 
should be the case in nature but for the sake of keeping the number of 
parameters under control we include this assumption. 
Thus a minimal set is $m_{{\tilde{e}}_L}^2$, $m_{{\tilde{e}}_R}^2$, $A_{\tau}$, 
$\varphi_{A_{\tau}}$.

For the squarks we can make similar assumptions so a minimal set is 
$m_{\tilde{Q}}^2$, $m_{\tilde{u}}^2$, $m_{\tilde{d}}^2$, $A_{b}$, $A_{t}$, 
$\varphi_{A_{b}}$, $\varphi_{A_{t}}$. To these we add from the Higgs sector 
$m_{h^0}$
(we have already counted $\tan\beta$), and the gravitino mass and phase 
$m_{\tilde{G}}$ and $\varphi_{\tilde{G}}$. Thus the MRM (minimal reasonable 
model) has 22 parameters. It could well happen that this is sufficient to 
describe the full Lagrangian. Small flavor changing effects could eventually be 
studied in rare decays or at colliders, but may not affect significantly the 
non-flavor-changing processes. This set of 22 MRM parameters is also 
self-contained (and separates into self-contained subsets) under 
renormalization group running. The MRM set is sensible to use for analyzing 
data in order to measure the Lagrangian soft parameters, and therefore in 
detector design studies as well. Once there is data it is of course also 
sensible to check simpler hypotheses as well but there is a danger that 
unphysical results will be obtained because the equations relating the soft 
parameters to the masses are non-linear and errors on masses and cross sections 
will not be small. Our MRM set allows for effects of D-terms and extra 
neutralinos from additional $U(1)$ symmetries and Planck scale operators, while 
a smaller parameter set 
does not. Since any given process depends 
on a few of the parameters the MRM set is not as large as it seems. We are 
not arguing that the full soft parameter set reduces to the MRM one, but 
that the MRM one is a reasonable place to start analyses that might be 
general. It could well be that different approaches to the flavor problem 
(such as horizontal symmetries) would lead to useful alternative MRM2, 
MRM3, {\it etc.}

In the MRM, one could then analyze LEP chargino data with only one extra 
parameter, the sneutrino mass. However, the normal destructive interference of 
$\gamma +Z$ and $\tilde{\nu}$ exchanges can be modified by the phase structure 
at the gauge boson-chargino vertices. More generally, it may be possible to 
determine enough observables at LHC (possibly even at FNAL) to measure the full 
set of MRM parameters.

\section{Concluding Remarks}

Once candidates for fundamental theory, e.g. string theories, are well enough 
understood to make contact with experiment some of their main predictions will 
be the 107 parameters of the Supersymmetry Lagrangian. Similarly, any 
theory of supersymmetry breaking will make such predictions. Measuring the 
parameters of the Lagrangian will be essential to test these predictions. If 
history is any guide, measurement of the Lagrangian parameters may be essential 
to lead the theorists to the correct theory of supersymmetry breaking or the 
correct string vacuum.

As supersymmetric theories were studied, it was appropriate to initially make 
many simplifying assumptions to gain intuition about the phenomena the theory 
predicted. Some of the usual assumptions may be true, some not. We have shown 
in this paper that observables depend more strongly on assumptions, 
particularly about phases, than has generally been realized. We have 
illustrated the 
effects of phases partly with the phase $\varphi_{\mu}$ of the $\mu$ parameter; 
$\varphi_{\mu}$ may or may not be large, but it should be interpreted here as a 
typical
one of a number of phases, most of which are only weakly constrained if at all.

In particular we have demonstrated that LEP and FNAL may have to be content 
with discovering the first superpartners (and a Higgs boson) and measuring some 
of their masses. Rigorous measurement of the Lagrangian parameters and 
$\tan\beta$ is not possible at these colliders (data on the masses cannot be 
uniquely converted into Lagrangian parameters). Before saying ``so what'', keep 
in mind that this would be somewhat analogous to a knowledge of the Standard 
Model with a few events of $Z$ production but no $W$ production and no 
measurement of 
$\sin^2\theta_W$ in different processes, no top quark, no study of $Z$ decays 
or of QCD jets, and no tests of the Higgs sector, not only no Higgs boson but 
no $\rho$ parameter either. We have also shown that a sufficiently energetic 
lepton collider with a polarized beam will be able to make the essential 
measurements needed to relate experiment to theory, and together with hadron 
colliders may be able to determine all of the Lagrangian parameters. We have 
defined a minimal set of parameters that allows for general effects (MRM); the 
MRM parameters may be measurable at LHC or possibly even FNAL.

Because of the sensitivities to parameters, limits on superpartner masses may 
be weaker than has been reported and need to be re-evaluated. This is also true 
for the Higgs mass limits, where the effect can be large.
  
\section*{Acknowledgments}

We appreciate helpful conversations with H. Haber, S. Martin and particularly 
with J. Wells (whom we also thank for emphasizing how large the effect on Higgs 
mass limit can be). This 
research was supported in part by the U.S. Department of Energy.

%

%
%

\end{document}